\documentclass[letter,12pt]{article}
\usepackage[top=1in,bottom=1in,left=1in,right=1in]{geometry}

\pdfoutput=1

\usepackage{amsmath,amssymb,amsfonts}
\usepackage{amsthm}         
\usepackage{latexsym}
\usepackage[scr=boondox]{mathalfa}
\usepackage{dsfont}
\usepackage{physics}
\usepackage{slashed}
\usepackage{simpler-wick}
\usepackage{cancel}
\usepackage{yhmath}
\usepackage{mathdots}
\usepackage{cite}
\usepackage[normalem]{ulem}

\usepackage{pgfplots}
\pgfplotsset{compat=1.17}
\usepackage{tikz}
\usepackage{tikz-3dplot}
\usetikzlibrary{
  3d,
  calc,
  fadings,
  patterns,
  decorations.pathreplacing,
  shadows.blur,
  shapes
}

\usepackage{booktabs}
\usepackage{multirow}
\usepackage{tabularx}
\usepackage{array}

\usepackage{gensymb}
\usepackage{siunitx}
\usepackage{pifont}

\usepackage[utf8]{inputenc}
\usepackage{verbatim}

\usepackage{hyperref}

\usepackage{caption}

\numberwithin{equation}{section}

\newcommand{\be}{\begin{equation}}
\newcommand{\ee}{\end{equation}}
\newcommand{\bs}{\begin{subequations}}
\newcommand{\es}{\end{subequations}}

\newcommand{\pa}{\partial}

\def\ve{{\varepsilon}}
\def\s{{\sigma}}

\def\a{{\alpha}}

\def\G{{\Gamma}}
\def\Th{{\Theta}}
\def\d{{\delta}}
\def\ka{{\kappa}}
\def\th{{\theta}}
\def\D{\Delta}
\def\O{\Omega}

\def\th{\theta}

\def\CB{{\mathcal B}}
\def\CC{{\mathcal C}}

\def\CE{{\mathcal E}}

\def\CH{{\mathcal H}}

\def\CL{{\mathcal L}}

\def\CO{{\mathcal O}}
\def\CP{{\mathcal P}}

\def\la{\langle}
\def\ra{\rangle}
\newcommand{\dt}{{\text d}}

\def\+{{(+)}}
\def\-{{(-)}}
\def\0{{(0)}}
\def\1{{(1)}}
\def\2{{(2)}}
\def\3{{(3)}}

\def\p{{\partial}}

\def\tr{{\tilde{r}}}

\def\rd{\text{d}}

\def\bth{{\bar\theta}}
\def\mid{{\text{mid}}}

\begin{document}
\begin{titlepage}
\unitlength = 1mm
\hfill CALT-TH 2025-009
\ \\
\vskip 2cm
\begin{center}

\openup .5em

{\Large{Quantum Area Fluctuations from Gravitational Phase Space}}

\vspace{0.8cm}
Luca Ciambelli,$^1$ Temple He,$^2$ Kathryn M. Zurek$^2$

\vspace{1cm}

{\it  $^1$Perimeter Institute for Theoretical Physics \\ 31 Caroline Street North, Waterloo, Ontario, Canada N2L 2Y5} \\
{\it  $^2$Walter Burke Institute for Theoretical Physics \\ California Institute of Technology, Pasadena, CA 91125 USA}\\

\vspace{0.8cm}

\begin{abstract}
 
We study the gravitational phase space associated to a stretched horizon within a finite-sized causal diamond in $(d+2)$-dimensional spacetimes. By imposing the Raychaudhuri equation, we obtain its constrained symplectic form using the covariant phase space formalism and derive the relevant quantum commutators by inverting the symplectic form and quantizing. Finally, we compute the area fluctuations of the causal diamond by taking a Carrollian limit of the stretched horizon in pure Minkowski spacetime, and derive the relationship $\langle (\Delta A)^2 \rangle \geq \frac{2\pi G}{d}\langle A \rangle$, showing that the variance of the area fluctuations is proportional to the area itself.
 \end{abstract}

\vspace{1.0cm}
\end{center}
\end{titlepage}
\pagestyle{empty}
\pagestyle{plain}
\pagenumbering{arabic}

\tableofcontents

\section{Introduction}

The physics at black hole horizons has long been a subject of interest and investigation in theoretical physics. Because of the null nature of the event horizon, it can often be difficult to understand aspects of its geometry and how fields propagate on it. However, one particularly useful tool that has yielded enormous progress in this subject is the black hole membrane paradigm, which introduces the concept of a stretched horizon, and was originally formulated and studied in \cite{Damour:1979wya, Price:1986yy, Thorne:1986iy, Susskind:1993if, Parikh:1997ma}. More recently, it has been used in the context of holography to study the hydrodynamic behavior of gravitational fluctuations \cite{Kovtun:2003wp, Iqbal:2008by, Starinets:2008fb, Bredberg:2010ky, Faulkner:2010jy}. 

While black hole horizons have garnered much attention, finite horizons in empty spacetime created by null rays, which is the subject of this paper, have been studied far less extensively. Nevertheless, in recent years, they are beginning to receive more attention, appearing as a central ingredient in the study of Ryu-Takayanagi entanglement wedges in the context of AdS/CFT \cite{Ryu:2006bv, DeBoer:2018kvc, Verlinde:2019ade}, gravitational thermodynamics in various backgrounds \cite{Jacobson:2015hqa,Jacobson:2018ahi,Verlinde:2019xfb}, shockwave geometries~\cite{Verlinde:1991iu, Verlinde:2022hhs, He:2023qha, He:2024vlp}, Carrollian geometry \cite{Penna:2015gza, Donnay:2019jiz, Ciambelli:2019lap, Chandrasekaran:2021hxc, Freidel:2022vjq, Ciambelli:2023mir, Ciambelli:2024swv, Freidel:2024emv}, null gravitational phase spaces \cite{
Reisenberger:2007pq, Parattu:2015gga, Lehner:2016vdi, Wieland:2019hkz, Adami:2020amw, Chandrasekaran:2023vzb, Odak:2023pga, Bub:2024nan, Ciambelli:2025mex}, inflationary cosmology \cite{Aalsma:2025bcg}, and conformal behavior of near-horizon geometries~\cite{Banks:2021jwj}.   
A commonality shared by all of the approaches mentioned above is that horizons created by null rays have properties in common with black hole horizons.  Furthermore, while the membrane paradigm has been extensively studied for black hole horizons, it has received limited attention within the context of null horizons~\cite{Bredberg:2011jq, Penna:2015gza, Penna:2017vms, Zhang:2023mkf,Bak:2024kzk}. 

\begin{figure}[t]
\begin{minipage}{0.45\textwidth}
\centering
\begin{tikzpicture}
    \draw[->] (0,-4) -- (0,4) node[above] {$t$};
    \draw[->] (-1,0) -- (4,0) node[right] {$r$};

    \draw[thick, black] (0,3) -- (3,0) -- (0,-3) --  cycle;
    \filldraw [cyan] (3,0) circle (2pt);
    \node at (3.3,0.3) {\small\textcolor{cyan}{$\CB$}};
    \node at (1,2.8) { $\CH^+$ };
    \node at (1,-2.8) { $\CH^-$ };
    \node at (1,-1) {\color{red} $\CH_s$ };

    \draw[thick, red, domain=1:3, samples=100, smooth] 
        plot (-{\x} + 3, {sqrt(\x^2 - 1)});
    \draw[thick, red, domain=1:3, samples=100, smooth] 
        plot (-{\x} + 3, {-sqrt(\x^2 - 1)});
        
    \draw[->, rotate around = {33:(1.71,0.81)}, thick] (1.71,0.81) -- (1.71,1.21) node[left] {\tiny $\tau$};
    \draw[->, rotate around = {-135:(1.71,0.81)}, thick] (1.71,0.81) -- (2.11,0.81) node[left] {\tiny $\rho$};
    
     \draw[thick] (2,-3) -- (3,-3);
    \draw[thick] (2,-2.9) -- (2,-3.1); 
    \draw[dashed] (2,-2.9) -- (2,0);
    \draw[thick] (3,-2.9) -- (3,-3.1);   
    \draw[dashed] (3,-2.9) -- (3,0);
    \node at (2.5,-3.5) { $\frac{ \sqrt{2h_0}}{\ka}$ };
    
     \draw[thick] (0,3.5) -- (3,3.5);
    \draw[thick] (0,3.4) -- (0,3.6); 
    \draw[dashed] (0,3.4) -- (0,3);
    \draw[thick] (3,3.4) -- (3,3.6);   
    \draw[dashed] (3,3.4) -- (3,0);
    \node at (1.5, 3.8) { $L$ };
    
     \draw[thick] (-0.1,3) -- (-0.1,2.8);
    \draw[thick] (-0.2,3) -- (0,3); 
    \draw[thick] (-0.2,2.8) -- (0,2.8);   
    \node at (-0.6, 2.9) { $\frac{h_0}{\ka^2 L}$ };
\end{tikzpicture}
\end{minipage}\hfill
\begin{minipage}{0.45\textwidth}
\centering
\tdplotsetmaincoords{85}{115} 
\begin{tikzpicture}[tdplot_main_coords, scale=2.8]
  \def\R{1} 
  \def\H{1} 

  \shade[cyan, thick, shading=radial, inner color=gray!40, outer color=gray!80] (0,0,0) circle (\R);
  \draw[thick,cyan] (0,0,0) circle (\R);
   
  \foreach \angle in {0,30,...,360} {
    \draw[] 
      (0,0,\H) -- ({\R*cos(\angle)}, {\R*sin(\angle)}, 0);
    \draw[]
      (0,0,-\H) -- ({\R*cos(\angle)}, {\R*sin(\angle)}, 0);
  }

  \node at (0,-1.2,0.3) {\small\textcolor{cyan}{$A = \O_d L^d$}};
  \node at (0,-1.2,0) {\small\textcolor{cyan}{$\CB$}};
  \node at (0,0.5,0.8) { $\CH^+$ };
  \node at (0,0.5,-0.8) { $\CH^-$ };

\end{tikzpicture}
\end{minipage}
\caption{On the left, we have a causal diamond of size $L$ in $d+2$ dimensions with $d$ angular directions suppressed. We depict the stretched horizon in red, which corresponds to a Rindler trajectory with acceleration $a = \frac{\ka}{\sqrt{2h_0}}$. The vector $\p_\tau$ is tangent to the stretched horizon, while the vector $\p_\rho$ is always null and parallel to the past horizon (it is not the normal to $\CH_s$ as $\pa_\rho \cdot \pa_\tau=1$). The distance between the stretched horizon and the bifurcating surface is given by $a^{-1}=\frac{\sqrt{2h_0}}{\ka}$, while the separation between the stretched horizon and the top and bottom tips of the causal diamond is $\frac{h_0}{\ka^2 L}$. On the right, we have drawn a 3-dimensional rendering of the causal diamond, with the bifurcate horizon $\CB$ (the intersection of $\CH^+$ and $\CH^-$) shown in blue. The area of the bifurcate horizon is $A = \O_d L^d$, where $\O_d$ is the surface area of a unit $d$-sphere.} 
\label{fig:sh}
\end{figure}
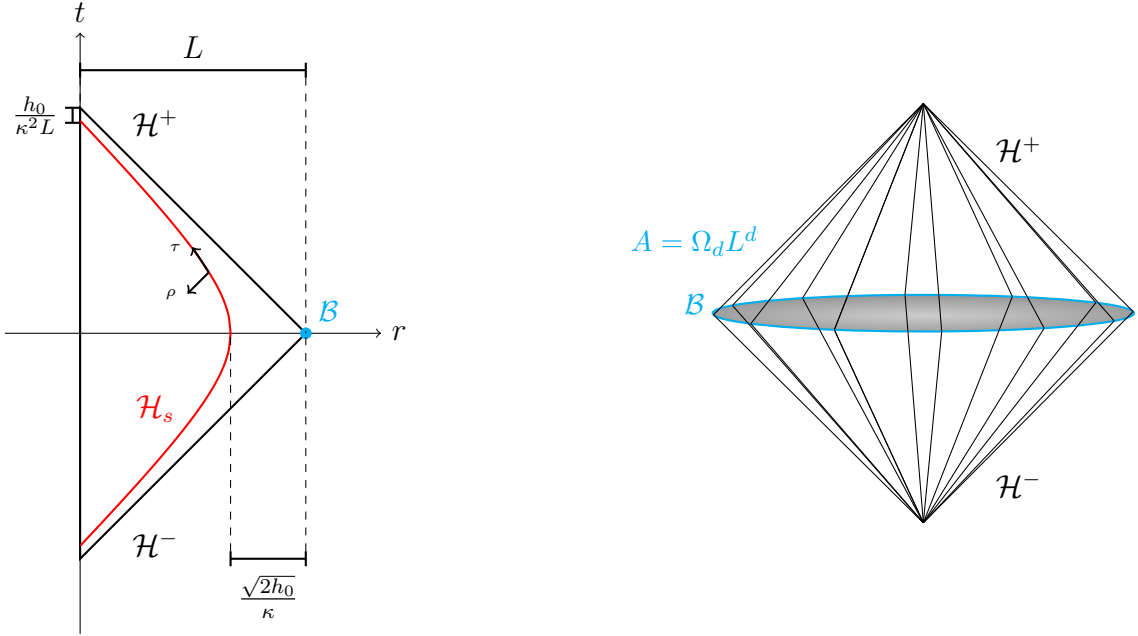

In this paper, we rigorously characterize the covariant phase space of a stretched horizon in a $(d+2)$-dimensional causal diamond on a background satisfying the Raychaudhuri equation. Since we are ultimately interested in the area fluctuations, we consistently decouple the spin-1 (twist) and spin-2 (radiative shear)  contributions.\footnote{We refer the reader to Appendix~\ref{AppA} for the precise restrictions on the field content performed here, and to \cite{Hopfmuller:2016scf, Chandrasekaran:2018aop, Hopfmuller:2018fni, Donnay:2019jiz, Odak:2023pga, Chandrasekaran:2023vzb, Ciambelli:2023mir} for the spin-0, spin-1, and spin-2 decomposition of the gravitational phase space.} 
We will thus be able to obtain the classical degrees of freedom on a stretched horizon associated to null rays.  Our goal is to  quantize these degrees of freedom so that we can determine how spacetime fluctuates near the null horizon of an empty causal diamond (see Figure~\ref{fig:sh}). This allows to better understand aspects of earlier proposals which characterize the quantization of null horizons via so-called modular fluctuations~\cite{Verlinde:2019xfb,Verlinde:2019ade,Banks:2021jwj,Verlinde:2022hhs,He:2024vlp,Bub:2024nan}, with the area operator we obtain from the covariant phase space analysis related to the modular operator.

Explicitly, we derive the constrained symplectic form in Section~\ref{ssec:raychau}, and the corresponding Dirac bracket on a fixed stretched horizon $\CH_s$ (shown in Figure~\ref{fig:sh}) is given in \eqref{bracket-c} to be 
\begin{align}\label{eq:intro-1}
\begin{split}
	\big\{\varphi(\tau,\vartheta) ,  \psi(\tau',\vartheta') \big\} = -  \frac{4\pi G}{h_0} \delta^{(d)}(\vartheta-\vartheta') H(\tau'-\tau) \pa_\tau\varphi(\tau,\vartheta)\,,
\end{split}
\end{align}
where the conjugate variables are $\varphi$, the size of the metric components transverse to a null ray (equivalent to the area of a sphere in the spherically symmetric limit), and $\psi = \p_\rho\varphi$, the change in $\varphi$ when moving away from $\CH_s$.  The coordinates on the stretched horizon are the clock $\tau$ and the angular coordinates $\vartheta$ transverse to the null ray. Furthermore, the location of $\CH_s$ relative to the null horizon is characterized by the dimensionless number $h_0 \equiv \kappa \rho_0$, with $h_0 \to 0$ being the limit where the stretched horizon $\CH_s$ reduces to the past and future null horizons $\CH^+ \cup \CH^-$ shown in Figure~\ref{fig:sh}. As we note in more detail, $\kappa$ is a convenient dimensionful constant that can be identified with the degree to which the clock $\tau$ fails to be affine; it can also be associated with the temperature of a Rindler observer (not necessarily the one on $\CH_s$).  Furthermore, because the $h_0 \rightarrow 0$ limit localizes $\CH_s$ onto null hypersurfaces, this is precisely when the theory becomes Carrollian \cite{Donnay:2019jiz}. Crucially, the $h_0 \to 0$ limit is singular in \eqref{eq:intro-1}, suggesting that to recover the phase space of $\CH^+ \cup \CH^-$ is not as straightforward as simply taking this limit.  

Next, we quantize the theory by promoting the Dirac bracket  in \eqref{eq:intro-1} to a quantum commutator. Confining our attention to Minkowski spacetime and performing an angle- and time-averaging of $\varphi$ over the entire stretched horizon, we obtain an uncertainty relation involving the averaged quantities. On the Minkowski background, we then relate \emph{both} the uncertainties of $\varphi$ and $\psi = \p_\rho\varphi$ to that of the transverse area of the causal diamond and can now also smoothly take the Carrollian limit $h_0 \to 0$. The final uncertainty relation is thus given in \eqref{final-result} to be
\be\label{eq:intro-2}
	\big\la (\D A)^2  \big\ra \geq \frac{2\pi G}{d} \la A \ra + \CO(h_0) \,.
\ee

We see that the size of the area fluctuations of the causal diamond containing $\CH_s$ in Minkowski spacetime has a lower bound not only involving the UV scale given by Newton's constant $G$, but also the IR scale given by $\la A \ra$. Because the $h_0 \to 0$ limit is perfectly smooth in \eqref{eq:intro-2}, this suggests that this equation holds even when $h_0 = 0$, corresponding to when our stretched horizon becomes the null boundaries of our causal diamond. In particular, we view this result as a concrete demonstration that it is possible for the IR scale to enter into the size of semiclassical metric fluctuations. It would be interesting to explore this observation further, and also to connect \eqref{eq:intro-2} to observables, in the future.  

Note that these area fluctuations are qualitatively similar to the fluctuations of the so-called modular Hamiltonian $K$, obtained either in the context of AdS/CFT \cite{Nakaguchi:2016zqi, DeBoer:2018kvc, Verlinde:2019ade} or Minkowski space~\cite{Banks:2021jwj,Verlinde:2022hhs,He:2024vlp}.  However, \eqref{eq:intro-2} is dependent on the transverse spacetime dimension $d$, whereas these earlier works analyzing the modular Hamiltonian found the dimension-independent inequality $\langle \Delta K^2 \rangle \geq \langle K \rangle$. While it may be tempting to exactly identify the modular Hamiltonian $K$ with the area $A$, as done in \cite{Wall:2011hj, Jafferis:2015del}, we remark that the modular fluctuations studied in the above references correspond to variations in temperature while keeping the causal diamond size fixed, whereas we are explicitly allowing the causal diamond size to fluctuate. Therefore, while we expect there to be a relationship between the fluctuations of $K$ and those of $A$ studied here, the precise relationship remains to be determined, a subject of future work.

The organization of this paper is as follows. In Section~\ref{sec:prelim}, we introduce coordinate conventions and derive the Raychaudhuri constraint. In Section~\ref{sec:phase-space}, we use the covariant phase space formalism to derive the symplectic form, and then from it obtain the Dirac brackets. We also quantize the phase space to obtain the Hilbert space associated to the stretched horizon in a background satisfying the vacuum Einstein's equation with no spin-1 or spin-2 data. In Section~\ref{sec:uncertain}, we derive a lower bound for the size of area fluctuations in Minkowski spacetime. We conclude with some discussion and future direction in Section~\ref{sec:discuss}.

\section{Preliminaries}\label{sec:prelim}

In this section, we will choose appropriate coordinates to study a stretched horizon in $(d+2)$-dimensional spacetime. We begin by constructing Gaussian null coordinates in Section~\ref{ssec:sphere}, which is particularly useful for studying causal diamonds in Minkowski spacetime. We will then generalize these coordinates in Section~\ref{ssec:raychau} to derive the Raychaudhuri equation in more general spacetimes.

\subsection{Gaussian Null Coordinates with Spherical Symmetry}\label{ssec:sphere}

We begin by describing $(d+2)$-dimensional Minkowski spacetime in outgoing null coordinates, with line element 
\be\label{bondi-metric}
	\dt s^2 = -\dt u^2 - 2 \,\dt u\,\dt r + r^2 \dt \O_d^2 \,,
\ee
where $u = t-r$ is the outgoing null time, and $\dt \O^2_d$ is the transverse metric of a unit $d$-dimensional sphere. Restricting ourselves to a causal diamond of width $L$ centered at the origin, which in coordinates \eqref{bondi-metric} corresponds to $-L \leq u \leq  L -2r$, the past and future null boundaries $\CH^\pm$ of the diamond are respectively given by $u = - L$ and $u + 2 r = L$. The bifurcate horizon $\CB$ is then located at the intersection of these two  boundaries, namely $u = -L$ and $r = L$, such that $t = 0$. 

We now review the diffeomorphism used to transform \eqref{bondi-metric} to Gaussian null coordinates (e.g., see \cite{Bub:2024nan}). Consider the coordinate transform from $(u,r)$ to $(\tau,\tilde r)$ coordinates, where
\be
\label{diffeo}
	u = L  - 2\Phi_0(\tau) \,, \qquad r = \Phi(\tau,\tr) \,.
\ee
A diffeomorphism of this form maps constant $u$ hypersurfaces to constant $\tau$ hypersurfaces. We want $\tau$ to increase to the future and $\tr$ to increase inwards, so that $\p_\tau \Phi_0 < 0$ and $\p_{\tr} \Phi < 0$. As was argued in \cite{Bub:2024nan}, defining the past horizon $\CH^-$ to be at $\tau = \tau_0$, we have the relations\footnote{These relations remain unchanged even though we have chosen to center our causal diamond slightly differently than that considered in \cite{Bub:2024nan}.}
\be\label{diffeo2}
	\Phi_0(\tau_0) = L\,, \qquad \lim_{\tr \to 0} \Phi(\tau,\tr) = \Phi_0(\tau) \,,
\ee
and the metric describing the interior of the causal diamond after gauge fixing is
\begin{align}\label{metric-cd}
\begin{split}
	\dt s^2 &= -2 \ka \tr \, \dt \tau^2 + 2 \,\dt \tau\,\dt \tr + \Phi(\tau,\tr)^{2} \dt \O_d^2 \\
	\Phi(\tau,\tr) &=  L - \frac{1}{2\ka} e^{\ka \tau + \a} - \tr e^{-\ka \tau - \a}  \,,
\end{split}
\end{align}
where $\ka$ is a spacetime constant (but not necessarily a phase space constant), and $\Phi$ reduces to $L$ at the bifurcate horizon, but more generally behaves like a dilaton. This is the standard line element in Gaussian null coordinates, with $\tr=0$ describing the future horizon $\CH^+$.\footnote{Another coordinate system oftentimes used to describe Minkowski spacetime is the topological black hole metric, which is given by $\dt s^2 = - \frac{\tr}{L} \dt t^2 + \frac{L}{\tr}\dt \tr^2 + \Phi^2\dt \O_d^2$, e.g., see \cite{Lee:2023kry}. We can obtain \eqref{metric-cd} from the topological black hole metric via the coordinate transform $t = \tau + L \log \frac{L}{\tr}$ if we identify $\ka = \frac{1}{2L}$.\label{fn:top}} 

We would now like to consider a stretched horizon $\CH_s$ rather than the null horizon, as shown in red in the left diagram in Figure~\ref{fig:sh}. Such stretched horizons have been studied extensively in the literature, either in the neighborhood of black hole horizons (e.g., see \cite{Damour:1979wya, Price:1986yy, Thorne:1986iy, Susskind:1993if, Parikh:1997ma, Iqbal:2008by, Starinets:2008fb, Bredberg:2010ky,Donnay:2019jiz}) or, more recently, near generic null horizons (e.g., see \cite{Bredberg:2011jq, Penna:2015gza, Chandrasekaran:2021hxc, Freidel:2022vjq, Freidel:2024emv}). To this end, it is convenient to define
\be\label{eq:rho-def}
	\tr = \rho + \rho_0 \quad\implies\quad  \dt \tr =  \dt \rho \,,
\ee
so that $\rho=0$ corresponds to the stretched horizon, and $\rho_0$ parametrizes its location with respect to $\CH^\pm$. Further defining $\varphi \equiv \Phi^d$, the metric becomes
\begin{align}\label{eq:gaussian_null_metric0}
\begin{split}
	\dt s^2 &= 2\, \rd \tau(\rd\rho-\ka ( \rho_0 + \rho) \dt \tau)+ \varphi(\tau,\rho)^{\frac{2}{d}} \dt \O_d^2 \\
	\varphi(\tau,\rho) &= \bigg( L - \frac{1}{2\ka} e^{\ka \tau + \a} - (\rho + \rho_0)e^{-\ka\tau-\a} \bigg)^d \,.
\end{split}
\end{align}
Notice that since \eqref{eq:gaussian_null_metric0} is diffeomorphic to Minkowski spacetime, it is a solution to the vacuum Einstein's equation and therefore in particular also satisfies the Raychaudhuri equation. As was argued in \cite{Bub:2024nan}, $\tau$ is the proper time for a uniformly accelerating observer at $\tilde r = \frac{1}{2 \kappa}$, with proper acceleration $\ka$ and Unruh temperature $T = \frac{\kappa}{2 \pi}$.  By further defining the dimensionless variable
\be\label{eq:h0-def}
	h_0=\kappa \rho_0 \,,
\ee
the line element \eqref{eq:gaussian_null_metric0} can be trivially rewritten as
\be\label{eq:gaussian_null_metric}
	\dt s^2 = 2\,\rd \tau(\rd\rho-(h_0+\ka\rho) \dt \tau)+ \varphi(\tau,\rho)^{\frac{2}{d}} \dt \O_d^2 \,.
\ee
The value of $\varphi$ on the stretched horizon has either been referred to as the breathing mode \cite{Binetruy:1993sj, Seraj:2021qja, Li:2022mvy, Bub:2023bfi, Lee:2023kry, Vermeulen:2024vgl}, or as the local area element in the context of null hypersurfaces.\footnote{Many previous papers \cite{Reisenberger:2007pq, Wieland:2017zkf, Hopfmuller:2018fni, Chandrasekaran:2018aop, Adami:2021nnf, Ciambelli:2023mir, Ciambelli:2024swv} denote $\varphi$ as $\O$. We will follow here instead the notation in \cite{Bub:2024nan}, as we use $\O$ to denote the symplectic form below.}  Furthermore, the quantity $\ka$ is precisely the inaffinity, which  vanishes when $\ell \equiv \p_\tau$ becomes affine.\footnote{More precisely, the inaffinity $\ka$ is defined by $\ell^\mu\nabla_\mu \ell^\nu\stackrel{\CH_s}{=}\kappa \ell^\nu$, where the equality is understood to hold when projected onto $\CH_s$. For $h_0=0$, $\ell$ is a conformal Killing vector of the induced metric \eqref{induced}, in which case the inaffinity coincides with the surface gravity on $\CH_s$, defined as $\nabla_\mu(\ell^\nu\ell_\nu)=-2\ka \ell_\mu$ (e.g., see \cite{Jacobson:1993pf}). Note that surface gravity is not uniquely defined and depends on the choice of clock. We also refer the reader to Footnote~\ref{fn:inaffinity} for a more physical understanding of why $\ka$ is the inaffinity.} As is apparent from \eqref{eq:gaussian_null_metric}, the inaffinity is captured by the subleading term in $\rho$ of $g_{\tau\tau}$.

In these coordinates, the induced metric on the stretched horizon $\CH_s$ is obtained by setting $\rho=0$, so that
\be\label{induced}
 	\dt s^2\big|_{\CH_s} = -2h_0\, \dt \tau^2  + \varphi(\tau,0)^{\frac{2}{d}} \dt \O_d^2 \, .
\ee
We remark that in addition to specifying the location of the stretched horizon, $h_0$ in \eqref{induced} also controls the norm of the clock on it. Indeed, time is generated on $\CH_s$ by $\ell$, which has norm
\be\label{ell}
	\ell^2=-2 h_0  \,.
\ee
The limit $h_0\to 0$ is the limit where the stretched horizon hugs the null horizon, where we retrieve the well-known horizon description \cite{Hopfmuller:2018fni, Chandrasekaran:2018aop, Adami:2021nnf, Ciambelli:2023mir, Ciambelli:2024swv}. It is clear from \eqref{induced} that $\sqrt{2h_0}$ can be identified with the speed of light, and so $h_0\to 0$ is a Carrollian limit, as described in \cite{Donnay:2019jiz}. In this case, the clock becomes the Carrollian vector defining the null geometry of the horizon. 

To determine the separation between the stretched horizon and the bifurcate horizon, we compute that the stretched horizon intersects the $t = 0$ hypersurface at $\tau = \tau_\mid$, where 
\be\label{Hs-mid}
	e^{\ka\tau_\mid+\a} = \sqrt{2h_0} \,,
\ee
which in turn implies
\be
	r = L -\frac{ \sqrt{2h_0}}{\ka} \,.
\ee
As the bifurcate horizon is located at $(t,r) = (0,L)$, it follows that the separation between the stretched horizon and the bifurcate horizon is $\frac{ \sqrt{2h_0}}{\ka}$.\footnote{In the topological black hole metric where $\ka = \frac{1}{2L}$ (see Footnote~\ref{fn:top}), this separation becomes $2\sqrt{\rho_0 L}$.} Notice that because the stretched horizon consists of (spherical) Rindler trajectories of an accelerating observer with acceleration $a$, and the separation between the Rindler trajectory and the causal horizons is given by $a^{-1}$, the Rindler trajectories making up the stretched horizon have acceleration and Unruh temperature
\be\label{eq:accel}
	a = \frac{\ka}{\sqrt{2h_0}} \,, \qquad T = \frac{\ka}{2\pi\sqrt{2h_0}} \,.
\ee
We have drawn the corresponding spacetime diagram illustrating this in the left diagram of Figure~\ref{fig:sh}.\footnote{As stated below \eqref{eq:gaussian_null_metric0}, $\tau$ is the proper time of the Rindler trajectory with $a = \ka$. Thus, unless $h_0 = \frac{1}{2}$, $\tau$ is \emph{not} the proper time along $\CH_s$.} This allows us to make contact with the Unruh effect \cite{Unruh:1976db, Unruh:1983ms} and the Bisognano-Wichmann study of geometric modular flows \cite{Bisognano:1976za} (e.g., see \cite{Sorce:2024zme, Caminiti:2025hjq} and references therein). 

We conclude this subsection by observing that while $\ell$ is tangent to $\CH_s$, the vector $\p_\rho$ is not normal to $\CH_s$.\footnote{The normal one-form is $\dt\rho$. Using \eqref{eq:gaussian_null_metric}, the normal vector on $\CH_s$ is thus given by $k = \p_\tau + 2h_0 \p_\rho$.
} Rather, notice from \eqref{eq:gaussian_null_metric} that the norm of $\p_\rho$ vanishes, and so $\p_\rho$ is a null vector. Indeed, it is always parallel to the past horizon $\CH^-$ (see Figure~\ref{fig:sh}). To see this explicitly, we need to relate the coordinates we have adopted in \eqref{eq:gaussian_null_metric} with lightcone coordinates $u \equiv t-r$ and $v \equiv t+r$. Using \eqref{diffeo}, \eqref{metric-cd}, and \eqref{eq:rho-def}, this is given by
\be \label{eq:uv}
	u = - L + \frac{1}{\ka} e^{\ka \tau + \a} \,, \qquad v = u + 2r = L - 2(\rho+\rho_0) e^{-\ka\tau-\a}\,.
\ee
The corresponding vector fields $\p_\tau$ and $\p_\rho$ on the stretched horizon are then given by
\be\label{tau-rho-approx}
	\p_\tau = e^{\ka\tau+\a}\p_u + 2h_0e^{-\ka\tau-\a}\p_v \,, \qquad \p_\rho = -2e^{-\ka\tau-\a}\p_v \,,
\ee
where we used $h_0 = \ka \rho_0$ and have set $\rho=0$.\footnote{From \eqref{tau-rho-approx}, note that $\p_\tau = e^{\ka\tau+\a}\p_u$ when $h_0 = 0$. Therefore, it becomes clear why $\ka$ is the inaffinity, since $\ka$ is merely reparametrizing the affine time $u$ along the null horizon.\label{fn:inaffinity}} Thus, we see that $\p_\rho$ is proportional to $-\p_v$, which is precisely the inward pointing vector field parallel to $\CH^-$.

We will later be interested in the case where $h_0 \to 0$, which as we mentioned above corresponds to $\CH_s$ hugging $\CH^+ \cup \CH^-$. From \eqref{tau-rho-approx}, it may be tempting to neglect the $\CO(h_0)$ term and conclude that $\p_\tau$ is always parallel to $\p_u$. However, this is only true near $\CH^+$. To see why, recall that we computed that the midpoint of the bifurcate horizon is given by $\tau= \tau_\mid$, which satisfies \eqref{Hs-mid}. Substituting this equation into \eqref{tau-rho-approx}, we obtain
\be
\p_\tau\big|_{\tau=\tau_\mid} = \sqrt{2h_0} (\p_u + \p_v) \,, \qquad \p_\rho\big|_{\tau=\tau_\mid} = - \sqrt{\frac{2}{h_0}}\p_v \,,
\ee
which is expected since $\p_\tau \sim \p_t$ at the midpoint of $\CH_s$ (see Figure~\ref{fig:sh}). Indeed, for $\tau < \tau_\mid$, it is clear from \eqref{tau-rho-approx} that it is actually the $\p_v$ component in $\p_\tau$ that dominates in the small $h_0$ expansion, as expected, since the tangent vector along the bottom half of $\CH_s$ is parallel to $\p_v$. Thus, in order to approximate $\p_\tau \sim \p_u$ in a small $h_0$ expansion, we need to assume $e^{\ka\tau + \a}$ is $\CO(1)$, which only occurs on $\CH_s$ near $\CH^+$.

\subsection{Deriving the Raychaudhuri Constraint}
Thus far, we have shown how to explicitly map Minkowski metric in outgoing null coordinates to the metric \eqref{eq:gaussian_null_metric}, useful for describing a stretched horizon, by utilizing the coordinate transform performed in \cite{Bub:2024nan}. Taking this as the starting point, we now generalize the line element \eqref{eq:gaussian_null_metric} by relaxing both spherical symmetry as well as the time-independence of $\ka$, while keeping $h_0$ constant. We thus have\footnote{The simplifications needed to derive \eqref{gengau} from a generic stretched horizon line element  are discussed in Appendix~\ref{AppA}.}
\begin{align}\label{gengau}
\begin{split}
	\dt s^2 &= - 2 F(\tau,\rho,\vartheta) \dt \tau^2 + 2 \, \dt \tau \,\dt \rho + \varphi(\tau,\rho,\vartheta)^{\frac{2}{d}} \dt\O_d^2 \\
	F(\tau,\rho,\vartheta) &= h_0 + \ka(\tau,\vartheta)\rho + \CO(\rho^2) \\
	\varphi(\tau,\rho,\vartheta) &= \varphi(\tau,0,\vartheta) + \rho\p_\rho\varphi(\tau,0,\vartheta) + \CO(\rho^2) \,,
\end{split}
\end{align}
where $\vartheta$ denotes the angular components, and we have Taylor expanded the functions $F$ and $\varphi$ about $\rho = 0$, the location of the stretched horizon $\CH_s$. 

Let us define the expansion scalars, which play a crucial role in what follows, to be
\be\label{pizza}
	\th = \frac{\p_\tau\varphi}{\varphi}\bigg|_{\rho=0} = \p_\tau\log\varphi \big|_{\rho = 0} \,, \qquad \bar\theta = \frac{\p_\rho\varphi}{\varphi} \bigg|_{\rho = 0} = \p_\rho\log\varphi \big|_{\rho=0} \,.
\ee
We are interested in their evolution along the stretched horizon $\CH_s$. Contracting Einstein's tensor $\mathbb E_{\mu\nu}$ with the tangent and normal vectors to $\CH_s$, given respectively by $\ell = \p_\tau$ and $k = \p_\tau + 2h_0 \p_\rho$, we obtain the constraint equation
\be\label{CiaoBello}
	\CC \equiv \lim_{\rho\to 0}\ell^\mu k^\nu \mathbb{E}_{\mu\nu} = \mathbb E_{\tau\tau} + 2h_0 \mathbb E_{\tau\rho} \,.
\ee
A direct computation then yields\footnote{Note that \eqref{RayEq} holds for all $d>0$. For $d=0$, \eqref{eq:surface-tension} is ill-defined, and indeed the dilaton degenerates to unity by \eqref{eq:gaussian_null_metric0}. Thus we restrict ourselves to $d>0$, that is, to spacetimes with at least $3$ dimensions.} 
\be\label{RayEq}
	\CC = (\p_\tau + \th) \th - \mu\th + 2 h_0 \bigg( \p_\tau \bar\theta + \frac{\th\bar\th}{d} \bigg) \,,
\ee
where we introduced the quantity $\mu$, defined to be
\be\label{eq:surface-tension}
	\mu = \ka + \frac{d-1}{d} \th \,.
\ee
This combination, introduced in \cite{Hopfmuller:2016scf, Hopfmuller:2018fni}, plays a crucial role in the symplectic analysis, since it is related to the conjugate variable of the breathing mode $\varphi$, as we derive in \eqref{symp-form1}, cf. the analysis in \cite{Ciambelli:2023mir,Bub:2024nan}.

The specific component \eqref{CiaoBello} of Einstein's tensor is the celebrated Raychaudhuri equation \cite{PhysRev.98.1123, Sachs:1961zz, Landau:1975pou} in arbitrary $(d+2)$-dimensional spacetime, which governs the evolution of the expansion scalars along the stretched horizon.\footnote{In deriving \eqref{RayEq}, we have restricted ourselves to the spin-$0$ data (such that \eqref{sheshe} is zero) and required $h_0$ to be constant. The more general form of the Raychaudhuri constraint is relegated to Appendix~\ref{AppA}.} As we will see in the next section, the relevant degrees of freedom on $\CH_s$ will be $\th$ and $\bar\th$, and we will use \eqref{RayEq} to study vacuum solutions by imposing the constraint $\CC = 0$ on the gravitational phase space on $\CH_s$.

To determine the details of where the stretched horizon begins and ends, note that this is precisely when the $\rho=0$ surface becomes a caustic. This corresponds to the angular part of the metric vanishing, implying from \eqref{gengau} that the past and future locations of the caustic along the stretched horizon occurs at $\tau = \tau_\pm$, where 
\be\label{caustic}
	\varphi(\tau_\pm, 0 , \vartheta) = 0 \,.
\ee
Such $\tau_\pm$ must exist, since otherwise, $\rho=0$ would not correspond to a stretched horizon depicted in red in the left diagram of Figure~\ref{fig:sh}. The specific choice of $\tau_\pm$ depends on the functional form of $\varphi$. Once we choose $\varphi$, we can fix $\th$ using the Raychaudhuri constraint \eqref{RayEq}, allowing us to solve for $\p_\rho\varphi$, or equivalently $\bar\theta$, along $\CH_s$. In the specific case when $\varphi$ is given by \eqref{eq:gaussian_null_metric0}, we can solve for $\tau_\pm$ via
\begin{align}
\begin{split}\label{eq:phi-pm}
	&0 = \varphi(\tau_\pm,0) = \bigg( L - \frac{1}{2\ka} e^{\ka \tau_\pm + \a} - \rho_0 e^{-\ka\tau_\pm - \a} \bigg)^d 
	\\
	\implies\quad &  e^{\ka\tau_\pm + \a} = \ka \bigg( L \pm \sqrt{L^2 - \frac{2h_0}{\ka^2} } \bigg) \,,
\end{split}
\end{align}
where we used the definition of $h_0$ given in \eqref{eq:h0-def}. We can further use \eqref{diffeo}, \eqref{diffeo2}, and \eqref{metric-cd} to compute the Minkowski time $t_\pm$ at the tips of the stretched horizon to be
\be
	t_\pm = L  - 2\Phi_0(\tau_\pm) = \pm L \bigg( 1 - \frac{h_0}{\ka^2 L^2} \bigg) + \CO(h_0^2) \,,
\ee
where we also used $r=0$ at the tips of $\CH_s$. Thus, the separation between the tips of $\CH_s$ and the causal diamond is, to leading order in small $h_0$, given by $\frac{h_0}{\ka^2 L}$.\footnote{In the topological black hole metric where $\ka = \frac{1}{2L}$ (see Footnote~\ref{fn:top}), this separation becomes $2\rho_0$.} Of course, we can allow for more general (angle-dependent) $\varphi(\tau,0,\vartheta)$ as well, and the resulting metric \eqref{gengau} is a solution to the vacuum Einstein's equation only if the Raychaudhuri constraint \eqref{RayEq} (as well as the constraints arising from the other components of Einstein's equations) is satisfied.

\section{Gravitational Phase Space}\label{sec:phase-space}

In this section, we discuss the simplified gravitational phase space induced on a stretched horizon, focusing on the spin-$0$ sector only. In Section~\ref{ssec:kinematic}, we construct the kinematic phase space without imposing the Raychaudhuri constraint and derive the Poisson brackets. This phase space has been previously studied, for instance in \cite{Hopfmuller:2016scf, Hopfmuller:2018fni, Freidel:2022vjq, Freidel:2024emv}. Next, we extend beyond their analysis by imposing the Raychaudhuri constraint in Section~\ref{ssec:raychau}, which allows us to construct the constrained phase space and derive the Dirac brackets.

\subsection{Kinematic Phase Space}\label{ssec:kinematic}

In this subsection, we derive the symplectic form associated to the canonical phase space without imposing the Raychaudhuri constraint \eqref{RayEq}. As we will be constraining ourselves to the gravitational phase space on $\CH_s$, where $\rho=0$, we will adopt the notation
\be
	\varphi(\tau,\vartheta) \equiv \varphi(\tau,\rho=0,\vartheta) \,, \quad \th(\tau,\vartheta) \equiv \th(\tau,\rho=0,\vartheta)  \,, \quad \bar\th(\tau,\vartheta) \equiv \bar\th(\tau,\rho=0,\vartheta) \,,
\ee
so that we do not need to specify $\rho=0$ everywhere explicitly. Furthermore, to emphasize that $\partial_\rho\varphi$ on the stretched horizon is an independent degree of freedom from $\varphi$, we denote
\be
	\psi(\tau,\vartheta) \equiv \partial_\rho\varphi(\tau,\rho,\vartheta) \big|_{\rho=0} \, ,
\ee
so that 
\be
\label{meglio}
\bar\th(\tau,\vartheta)=\frac{\psi(\tau,\vartheta)}{\varphi(\tau,\vartheta)} \,.
\ee

To derive the pre-symplectic potential $\Theta$ in gravity on some hypersurface $\Sigma$, we vary the Einstein-Hilbert action and extract the boundary term (e.g., see \cite{Carroll:2004st}), and the result is
\be\label{pre-symp0}
	\Th = \frac{1}{16\pi G} \int_{\Sigma} \dt \Sigma_\mu \big( g^{\mu\rho} \nabla^\nu \delta g_{\nu\rho} - g^{\rho\nu}\nabla^\mu \delta g_{\rho\nu} \big) \,.
\ee
We want to consider the case where $\Sigma = \CH_s$, in which case the outward-pointing normal one-form is $-\rd \rho$, and the measure on $\CH_s$ is given by
\be
	\dt \Sigma_\mu\big|_{\CH_s} = - \delta^\rho_\mu\varphi(\tau,\vartheta)\,\dt \tau\, \dt \O_d \,,
\ee
where we have taken $\rho=0$, and $\dt\O_d$ implicitly contains the determinant factor in the transverse angular directions. Substituting the metric \eqref{gengau} into \eqref{pre-symp0}, we obtain
\be\label{thetacan}
\Th = - \frac{1}{8\pi G} \int_{\CH_s} \dt \tau\,\dt \O_d\, \bigg[ \bigg(\mu + \frac{2(d-1)}{d} h_0 \bar\theta \bigg) \delta\varphi + \psi \delta h_0 + \delta \big( 2 h_0 \psi + \ka \varphi + \p_\tau\varphi \big) \bigg] \,,
\ee
where $\mu$ is defined in \eqref{eq:surface-tension}.\footnote{In deriving \eqref{thetacan}, we only kept track of spin-0 contributions. More generally, there are also spin-1 and spin-2 contributions, and the gravitational phase space induced on a stretched horizon takes a rather complicated form. We review its derivation in Appendix~\ref{AppA}, together with the set of assumptions required to isolate the spin-$0$ contributions, thereby reproducing \eqref{thetacan}.}

To obtain the pre-symplectic form, we act with $\delta$ to obtain
\be\label{symp-form0}
	\O = \delta  \Th
	= - \frac{1}{8\pi G} \int_{\CH_s} \dt \tau\,\dt \O_d\, \bigg[ \d \bigg(\mu + \frac{2(d-1)}{d} h_0 \bar\theta \bigg) \wedge \delta\varphi + \delta\psi \wedge \delta h_0  \bigg] \,,
\ee
where we noted $\delta^2 = 0$, and so all the total variation terms in $\Th$ do not contribute to the pre-symplectic form. To proceed, it is reasonable to assume that $h_0$ is not only a spacetime constant, but a phase space constant as well, so that\footnote{The phase space with varying $h_0$ describes a family of stretched horizons, and is currently under development.}
\be\label{h0-constraint}
	\delta h_0 = 0 \,.
\ee
Physically, this assumption amounts to fixing ourselves onto a particular stretched horizon, and in this special case, we obtain upon substituting \eqref{h0-constraint} into \eqref{symp-form0}
\be\label{symp-form1}
	\O  = \frac{1}{8\pi G} \int_{\CH_s} \dt \tau\,\dt \O_d\, \d\varphi \wedge  \d \bigg(\mu + \frac{2(d-1)}{d} h_0 \bar\theta \bigg) \,,
\ee
where we used $\delta a \wedge \delta b = -\d b \wedge \d a$. Notice that $\O$ in \eqref{symp-form1} is now in Darboux form\footnote{By Darboux form, we mean the symplectic form can be written as $\d p \wedge \d q$ for a symplectic pair $(p,q)$.} and therefore manifestly invertible (ignoring the constraint \eqref{RayEq}), implying it is not merely the pre-symplectic form, but rather the symplectic form on the kinematic phase space. 

Indeed, inverting the symplectic form, we obtain the kinematic Poisson bracket
\be\label{poisson}
	\left\{ \varphi(\tau,\vartheta), \mu(\tau',\vartheta')+ \frac{2(d-1)}{d}h_0 \bar\theta(\tau',\vartheta')\right\} = -8\pi G \delta(\tau-\tau')\delta^{(d)}(\vartheta-\vartheta')\,.
\ee
Because of the restriction \eqref{h0-constraint}, this bracket is evaluated on the single stretched horizon $\CH_s$ located at $\rho=0$. For $h_0\to 0$, this bracket reduces exactly to the spin-$0$ sector of the null hypersurface phase space discussed in \cite{Ciambelli:2023mir,Bub:2024nan}. Therefore, the kinematic symplectic data are the breathing mode $\varphi$ and the linear combination $\mu + \frac{2(d-1)}{d}h_0 \bar\theta$. Note that the latter contains the scalar $\bar\theta$ measuring expansion away from $\CH_s$, which involves $\psi$ and is thus a new degree of freedom on $\CH_s$ independent of $\varphi$. Indeed, from the last line of \eqref{gengau}, it is clear $\psi=\p_\rho\varphi|_{\rho=0}$ is the subleading term in the expansion of $\varphi$ about $\rho=0$. This concludes our analysis of the spin-$0$ sector of the kinematic phase space induced by gravity on a stretched horizon. Next, we impose the Raychaudhuri constraint to construct the constrained phase space. 

\subsection{Imposing the Raychaudhuri Constraint}\label{ssec:raychau}

We would now like to impose the Raychaudhuri constraint \eqref{RayEq} in the symplectic form \eqref{symp-form1} to obtain the constrained symplectic form. Setting $\CC=0$ and solving for $\mu$, we obtain
\be\label{raych2}
	\mu = \th + 2h_0\frac{\bar\th}{d} + \frac{\p_\tau\th}{\th} + 2h_0 \frac{\p_\tau\bar\th}{\th} \,.
\ee
Substituting this into \eqref{symp-form1}, we get\footnote{If $\theta$ vanishes at a point, the Raychaudhuri constraint \eqref{RayEq} reduces to $h_0 \partial_\tau \bar\theta = 0$ there. Thus, \eqref{raych2} assumes $\theta$ is nowhere vanishing. However, if $\theta$ vanishes only on a measure-zero set, the integral in \eqref{symp-form1} remains unaffected, and the symplectic form is still given by \eqref{symp-int1}. We  henceforth make this assumption.}
\be\label{symp-int1}
	\O =  \frac{1}{8\pi G} \int_{\CH_s} \dt \tau\, \dt \O_d \,\delta\varphi \wedge \delta\bigg( \th + \frac{\p_\tau\th}{\th}  + 2h_0 \frac{\p_\tau\bar\th}{\th} + 2 h_0  \bar\theta \bigg) \,.
\ee
To simplify this expression, we observe that the first two terms in \eqref{symp-int1} combine to give
\begin{align}\label{simple-1}
\begin{split}
	\frac{1}{8\pi G} \int_{\CH_s} \dt \tau\, \dt \O_d \, \delta\varphi \wedge \delta\bigg( \th + \frac{\p_\tau\th}{\th} \bigg)  &= \frac{1}{8\pi G} \int_{\CH_s} \dt \tau\, \dt \O_d \, \delta\varphi \wedge \big( \d\th + \p_\tau\delta \log\th \big) \\
	&= \frac{1}{8\pi G} \int_{\CH_s} \dt \tau\, \dt \O_d \big( \delta \varphi \wedge \delta \theta - \p_\tau \delta\varphi \wedge \delta \log\th \big)  \\
	&= \frac{1}{8\pi G} \int_{\CH_s} \dt \tau\, \dt \O_d  \bigg( \delta \varphi \wedge \delta \theta - \th\delta\varphi \wedge \frac{\d\th}{\th} \bigg) \\
	&= 0 \,,
\end{split}
\end{align}
where in the second equality we integrated by parts over $\tau$ and dropped the boundary term, as the boundary of the stretched horizon intersects the $r=0$ caustic, implying $\d\varphi|_{\p\CH_s} = 0$; and in the third equality we used the variational form of \eqref{pizza} and the antisymmetry of the wedge product. Substituting this result back into \eqref{symp-int1}, we obtain
\begin{align}\label{symp-int2}
\begin{split}
	\O &=  \frac{h_0}{4\pi G} \int_{\CH_s} \dt \tau\, \dt \O_d \,\delta\varphi \wedge \d \bigg(\frac{\p_\tau\bar\th}{\th}  + \bar\th \bigg) \\
	&= \frac{h_0}{4\pi G} \int_{\CH_s} \dt \tau\, \dt \O_d \,\delta\varphi \wedge \d \bigg( \frac{\p_\tau \psi}{\p_\tau\varphi} \bigg) \\
	&= \frac{h_0}{4\pi G} \int_{\CH_s} \dt \tau\, \dt \O_d \,\delta\varphi \wedge \d \bigg( \frac{\p_\tau (\varphi\bar\th) }{\p_\tau\varphi} \bigg) \,,
\end{split}
\end{align}
where we repeatedly used \eqref{meglio} and the expression for $\theta$ in \eqref{pizza}. This is the final expression for the constrained symplectic form on the stretched horizon, and is a main result of this subsection. It is clear that the symplectic pair consists of the breathing mode $\varphi$ and the expansion away from the stretched horizon $\bar\th$. Moreover, we remark that aside from the trivial angular integral over a $d$-dimensional sphere, the symplectic form \eqref{symp-int2} is entirely independent of the spacetime dimension! All factors involving $d$ have dropped out in \eqref{symp-int1}, indicating that the physics of the spin-$0$ breathing mode is universal.

To derive the Dirac brackets along the stretched horizon, we simply invert the symplectic form \eqref{symp-int2} to obtain
\begin{align}\label{dirac-1}
\begin{split}
	\big\{ \varphi(\tau,\vartheta) , \varphi(\tau',\vartheta') \big\} &= 0 \\
	\bigg\{ \frac{\p_{\tau}(\varphi(\tau,\vartheta) \bar\th(\tau,\vartheta))}{\p_{\tau}\varphi(\tau,\vartheta)}  , \frac{\p_{\tau'}(\varphi(\tau',\vartheta') \bar\th(\tau',\vartheta'))}{\p_{\tau'}\varphi(\tau',\vartheta')} \bigg\} &= 0 \\ 
	\bigg\{\varphi(\tau, \vartheta) , \frac{\p_{\tau'}(\varphi(\tau',\vartheta') \bar\th(\tau',\vartheta'))}{\p_{\tau'}\varphi(\tau',\vartheta')} \bigg\} &= -\frac{4\pi G}{h_0}\delta(\tau-\tau')\delta^{(d)}(\vartheta-\vartheta') \,.
\end{split}
\end{align}
Notice that by taking a $\tau'$ derivative of the first line of \eqref{dirac-1}, we can immediately derive 
\be\label{bracket-2}
	\big\{ \varphi(\tau,\vartheta), \p_{\tau'}\varphi(\tau',\vartheta') \big\} = 0 \quad\implies\quad  \big\{ \varphi(\tau,\vartheta) , \th(\tau',\vartheta') \big\} = 0 \,,
\ee
where the implication follows from \eqref{pizza} and the fact the bracket between $\varphi$ and itself vanishes.\footnote{This should be contrasted with the situation for a null horizon, where the canonical momentum is $\theta$ itself, and thus the canonical momentum is not independent of $\varphi$ (e.g., see \cite{Ciambelli:2023mir} or the analysis carried out in \cite{Barnich:2024aln} for a scalar field).} Next, we will utilize \eqref{bracket-2} to derive the bracket between $\varphi$ and $\bar\th$, which we denote as
\be\label{eq:D-def}
	D(\tau,\vartheta,\tau',\vartheta') \equiv \big\{ \varphi(\tau,\vartheta) , \bar\th(\tau',\vartheta') \big\} \,.
\ee
Using this definition, the last line of \eqref{dirac-1} becomes the differential equation
\be\label{comm-ode}
	D(\tau,\vartheta,\tau',\vartheta') + \frac{1}{\th(\tau',\vartheta')}\p_{\tau'} D(\tau,\vartheta,\tau',\vartheta') =  -\frac{4\pi G}{h_0} \delta(\tau-\tau')\delta^{(d)}(\vartheta-\vartheta') \,,
\ee
where we used the fact the bracket between $\varphi$ and $\p_\tau\varphi$ vanish by \eqref{bracket-2}. This is a differential equation in the stretched horizon clock $\tau \in [\tau_-, \tau_+]$, where we recall from \eqref{caustic} that $\tau_\pm$ denote the endpoints of the stretched horizon $\CH_s$. The solution to \eqref{comm-ode} can be directly computed to be
\begin{align}\label{DD}
\begin{split}
	D(\tau,\vartheta,\tau',\vartheta') &= D(\tau,\vartheta,\tau_-,\vartheta') \exp\bigg[ - \int_{\tau_-}^{\tau'} \dt\tau''\, \th(\tau'',\vartheta') \bigg] \\
	&\qquad - \frac{4\pi G}{h_0} \delta^{(d)}(\vartheta-\vartheta') H(\tau' - \tau) \th(\tau,\vartheta') \exp\bigg[ - \int_{\tau}^{\tau'} \dt \tau''\, \th(\tau'',\vartheta') \bigg] \,, 
\end{split}
\end{align}
where $H(x)$ is the Heaviside step function defined to be $1$ for $x > 0$ and $0$ for $x\leq  0$. Now, the first term in \eqref{DD} is 
\begin{align}\label{exp-simp}
\begin{split}
	&D(\tau,\vartheta,\tau_-,\vartheta') \exp\bigg[ - \int_{\tau_-}^{\tau'} \dt\tau''\, \th(\tau'',\vartheta') \bigg] \\
	&\qquad= D(\tau,\vartheta,\tau_-,\vartheta') \exp\bigg[ - \int_{\tau_-}^{\tau'} \dt\tau''\, \p_{\tau''} \log \varphi (\tau'',\vartheta') \bigg]  \\
	&\qquad= D(\tau,\vartheta,\tau_-,\vartheta') \frac{\varphi(\tau_-,\vartheta')}{\varphi(\tau',\vartheta')} \\
	&\qquad= 0 \, ,
\end{split}
\end{align}
where we used both \eqref{pizza} and the caustic condition $\varphi(\tau_-,\vartheta') = 0$, namely \eqref{caustic}. It follows only the second term in \eqref{DD} contributes, and we finally have 
\begin{align}\label{bracket-b}
\begin{split}
	D(\tau,\vartheta,\tau',\vartheta') &= - \frac{4\pi G}{h_0} \delta^{(d)}(\vartheta-\vartheta') H(\tau' - \tau) \th(\tau,\vartheta') \exp\bigg[ - \int_{\tau}^{\tau'} \dt \tau''\, \th(\tau'',\vartheta') \bigg] \\
	&= - \frac{4\pi G}{h_0} \delta^{(d)}(\vartheta-\vartheta') H(\tau' - \tau) \frac{\p_\tau \varphi(\tau,\vartheta) }{\varphi(\tau',\vartheta')} \,,
\end{split}
\end{align}
where we used the same manipulations as those in \eqref{exp-simp} and then recalled \eqref{pizza}. 

As it turns out, it will later be more useful for us to write \eqref{bracket-b} in a slightly different but equivalent form. From \eqref{bracket-b}, we compute

\be\label{bracket-c}
	\big\{\varphi(\tau,\vartheta) ,  \psi(\tau',\vartheta') \big\} =-  \frac{4\pi G}{h_0} \delta^{(d)}(\vartheta-\vartheta') H(\tau'-\tau) \pa_\tau\varphi(\tau,\vartheta) \,,
\ee
where we used the definition of $\bth$ in \eqref{meglio}. We can now quantize by promoting the Dirac bracket in \eqref{uncertain1} to a quantum commutator via $[\cdot,\cdot] = i \{\cdot, \cdot\}$.\footnote{The quantization we perform here is on a timelike hypersurface and hence differs from the typical quantization on a spacelike hypersurface. It would be very interesting and relevant to understand how the two quantizations are related when $h_0 \to 0$, which is the limit we are ultimately interested in. We thank Prahar Mitra for discussions regarding this point.} This yields
\be\label{comm-c}
	\boxed{ \big[\varphi(\tau,\vartheta) ,  \psi(\tau',\vartheta') \big] =-  \frac{4\pi i G}{h_0} \delta^{(d)}(\vartheta-\vartheta') H(\tau'-\tau) \pa_\tau\varphi(\tau,\vartheta) \,, }
\ee
and is the main result of this subsection. 

We conclude by briefly remarking that even though we derived in \eqref{bracket-c} the Dirac brackets involving $\varphi$ and $\psi$, we could have equivalently also derived the brackets involving $\theta$ and $\bar\th$. Such brackets are not as useful for our purposes, but for completeness we have relegated their derivations to Appendix~\ref{app:grav-phase}.

\section{Deriving the Area Uncertainty} \label{sec:uncertain}

Having derived the commutator between $\varphi$ and $\psi$, we can now proceed to compute the Heisenberg uncertainty relation associated to the operators. In particular, we will be interested in deriving an area uncertainty associated to a causal diamond in $(d+2)$-dimensional Minkowski spacetime with length $L$. Because we are not interested in the angular components, we can integrate over both $\vartheta,\vartheta'$ and divide by the total solid angle $\O_d = \frac{2\pi^{\frac{d+1}{2} }}{\G(\frac{d+1}{2})}$. This can equivalently be thought of isolating the $S$-wave contribution in the bracket, and the result from averaging the angular components of \eqref{comm-c} is
\be\label{bracket-c2}
	\big[ \varphi_S(\tau) , \psi_S(\tau') \big] = -\frac{4\pi  iG}{h_0 \O_d}  H(\tau'-\tau) \p_\tau\varphi_S(\tau)\,,
\ee
where we defined
\be
	\varphi_S(\tau) \equiv \frac{1}{\O_d} \int_{S^d} \dt\O_d\, \varphi(\tau,\vartheta) \, , \qquad \psi_S(\tau) \equiv \frac{1}{\O_d} \int_{S^d} \dt\O_d\, \psi(\tau,\vartheta) \,.
\ee
Furthermore, we would like to perform a time-averaging of the bracket \eqref{bracket-c2}. Integrating \eqref{bracket-c2} over the entire range of the stretched horizon and then dividing by the total time duration $T \equiv \tau_+ - \tau_-$, we obtain
\begin{align}\label{integrate-bracket1}
\begin{split}
	\frac{1}{T} \int_{\tau_-}^{\tau_+} \dt \tau\, \big[ \varphi_S(\tau) ,  \psi_S(\tau') \big] &= - \frac{4\pi i G}{h_0 \O_d T} \int_{\tau_-}^{\tau_+} \dt\tau\, H(\tau'-\tau) \pa_\tau\varphi_S(\tau) \\
	 &= - \frac{4\pi i G}{h_0 \O_d T} \bigg[  H(\tau'-\tau) \varphi_S(\tau) \Big|_{\tau=\tau_-}^{\tau= \tau_+} + \int_{\tau_-}^{\tau_+} \dt\tau\, \d(\tau'-\tau) \varphi_S(\tau) \bigg] \\
	&= - \frac{4\pi i G}{h_0 \O_d T} \varphi_S(\tau') \,,
\end{split}
\end{align}
where we integrated by parts and then used the fact $\varphi(\tau_\pm,\vartheta)$, and therefore $\varphi_S(\tau_\pm)$, vanishes by \eqref{caustic}. Indeed, we can rewrite the above equality as
\be\label{bar-varphi}
	\big[ \bar\varphi_S ,  \psi_S(\tau') \big]  =  - \frac{4\pi i G}{h_0 \O_d T} \varphi_S(\tau') \,, \qquad \bar\varphi_S \equiv \frac{1}{T}\int_{\tau_-}^{\tau_+} \dt\tau \,\varphi_S(\tau) \,,
\ee
where $\bar\varphi_S$ is the time-averaged $S$-wave contribution in $\varphi$. Performing a further time-averaging over $\tau'$, we obtain the simplified commutator
\be\label{uncertain1}
	\big[ \bar\varphi _S, \bar\psi_S \big] = -\frac{4\pi i G}{h_0 \O_d T} \bar \varphi_S \,, \qquad \bar\psi_S \equiv \frac{1}{T}\int_{\tau_-}^{\tau_+} \dt\tau \,\psi_S(\tau) \,.
\ee
As $\bar\varphi_S$ and $\bar\psi_S$ do not commute, there exists a Heisenberg uncertainty relation between them, and it is given by
\be\label{uncertainty-1}
	\sqrt{\la (\D\bar\varphi_S)^2 \ra\la (\D\bar\psi_S)^2 \ra} \geq \frac{2\pi G}{h_0\O_d T} \la \bar\varphi_S \ra \,,
\ee
where the left-hand side of the equation is the product of the standard deviation of $\bar\varphi_S$ and $\bar\psi_S$ when inserted into some quantum state.\footnote{For any operator $\CO$, we define its variance to be $\la (\D \CO )^2 \ra \equiv \la \CO^2 \ra - \la \CO \ra^2$, and the standard deviation is the square root of the variance.}

Let us now consider the specific case where $\varphi$ and $\psi=\p_\rho\varphi|_{\rho=0}$ are given by \eqref{eq:gaussian_null_metric}, so that we are restricting ourselves to an isolated causal diamond in Minkowski spacetime. Furthermore, we will also take $h_0 \to 0$, so that $\CH_s$ hugs the null boundary $\CH^+ \cup \CH^-$. As we mentioned below \eqref{ell}, this is precisely the Carrollian limit for this theory where the stretched horizon maps smoothly onto the null horizon. In this case, note that we computed in \eqref{eq:phi-pm} the value of $\tau_\pm$, and the total time duration $T \equiv \tau_+ - \tau_-$ is to leading order in $h_0$
\be\label{T-compute}
	T = \frac{1}{\ka} \log \frac{2\ka^2 L^2}{h_0} + \CO(1) \,.
\ee
Performing the angle- and time-averaging of $\varphi$, we get using \eqref{eq:gaussian_null_metric0} and \eqref{eq:h0-def}
\be\label{int-varphiS}
	\frac{1}{T\Omega_d} \int_{\tau_-}^{\tau_+} \dt \tau \int_{S^d} \dt\O_d\,\varphi(\tau,\rho) 
	= \frac{1}{T} \int_{\tau_-}^{\tau_+} \dt \tau \bigg[ L - \frac{1}{2\ka} e^{\ka\tau+\a} - \bigg(\rho + \frac{h_0}{\ka} \bigg) e^{-\ka\tau-\a} \bigg]^d \,,
\ee
where we have carried out the angular integral trivially since $\varphi$ is spherically symmetric. We would now like to expand this to linear order in $\rho$ to extract $\bar\psi_S=\p_\rho\bar\varphi_S|_{\rho=0}$, while keeping only terms leading in a small $h_0$ expansion at each order in the small $\rho$ expansion. The result is given by
\begin{align}\label{int-1}
\begin{split}
	&\frac{1}{T\Omega_d} \int_{\tau_-}^{\tau_+} \dt \tau \int_{S^d} \dt\O_d\,\varphi(\tau,\rho) \\
	&\qquad= \frac{1}{T} \int_{\tau_-}^{\tau_+} \dt \tau  \bigg( L - \frac{1}{2\ka} e^{\ka\tau+\a}  - \rho e^{-\ka\tau-\a}\bigg)^d  + \cdots \\
	&\qquad= \frac{1}{T} \int_{\tau_-}^{\tau_+} \dt \tau  \bigg( L - \frac{1}{2\ka} e^{\ka\tau+\a} \bigg)^d \bigg(1 + \frac{2d \ka \rho}{e^{2(\ka\tau+\a)} - 2 \ka L e^{\ka\tau+\a}} \bigg)   + \cdots  \\
	&\qquad= L^d   + \frac{L^d}{\ka T}\sum_{k=1}^d \binom{d}{k} \frac{(-1)^k}{(2\ka L)^k k} \big( e^{k(\ka\tau_+ + \a)} - e^{k(\ka\tau_- + \a )} \big) \\
	&\qquad\qquad + \frac{2d\ka\rho L^d}{T} \Bigg[ \int_{\tau_-}^{\tau_+} \dt\tau \frac{1}{e^{2(\ka\tau+\a)} - 2\ka L e^{\ka\tau+\a}} \\
	&\qquad\qquad  + \sum_{k=1}^d \binom{d}{k} \frac{(-1)^k}{(2\ka L)^k } \int_{\tau_-}^{\tau_+} \dt\tau \frac{e^{k(\ka\tau+\a)}}{e^{2(\ka\tau+\a)} - 2\ka L e^{\ka\tau+\a}} \Bigg] + \cdots \,,
\end{split}
\end{align}
where $\cdots$ indicate terms that are subleading in the $h_0,\rho \to 0$ limit, and to obtain the final equality, we used the binomial theorem and separated out the $k=0$ case for later convenience. The first integral in the final equality of \eqref{int-1} can be computed to be
\begin{align}\label{int-1a}
\begin{split}
	&\int_{\tau_-}^{\tau_+} \dt\tau \frac{1}{e^{2(\ka\tau+\a)} - 2\ka L e^{\ka\tau+\a}} \\
	&\qquad= \bigg( \frac{1}{2\ka^2 Le^{\ka\tau+\a}} -\frac{\tau}{4 \ka^2 L^2} + \frac{1}{4\ka^3 L^2} \log\big( 2\ka L - e^{\ka\tau+\a} \big) \bigg)\bigg|_{\tau=\tau_-}^{\tau=\tau_+}  \\
	&\qquad=  - \frac{1}{2\ka h_0}  +  \cdots ,
\end{split}
\end{align}
where we used \eqref{eq:phi-pm} and then kept only the leading divergent term in the small $h_0$ limit. The final integral in \eqref{int-1} is more involved and can be expressed in terms of the incomplete beta function. However, as we are only interested in the leading divergent behavior in the $h_0 \to 0$ limit, we can ignore the final integral in \eqref{int-1} since it is subleading to \eqref{int-1a} in the $h_0 \to 0$ limit. Substituting these results into \eqref{int-1}, we see that
\begin{align}\label{time-avg-final}
\begin{split}
	&\frac{1}{T\Omega_d} \int_{\tau_-}^{\tau_+} \dt \tau \int_{S^d} \dt\O_d\,\varphi(\tau,\rho) \\
	&\qquad= L^d + \frac{1}{T}\sum_{k=1}^d \binom{d}{k} \frac{(-1)^kL^{d-k}}{(2\ka)^k \ka k}   ( e^{k(\ka\tau_+ + \a)} - e^{k(\ka\tau_- + \a )} ) -  \frac{\rho d L^d }{h_0 T} +\cdots   \\
	&\qquad=  L^{d} + \frac{L^d}{\ka T}\sum_{k=1}^d \binom{d}{k} \frac{(-1)^k}{k}  - \frac{\rho d L^d }{h_0 T}  +\cdots   \\
	&\qquad= L^{d} - \frac{\rho d L^d }{h_0 T}  +\cdots \,,
\end{split}
\end{align}
where in the second equality we used \eqref{eq:phi-pm} expanded about $h_0$, and in the last equality we used $T \sim \log h_0$ to the conclude that the second term is subleading. We remind the reader that in the above expression, we are first performing a small $\rho$ expansion to linear order, and then consistently keeping only leading terms in a small $h_0$ expansion at each order in $\rho$. Using \eqref{time-avg-final} on the stretched horizon, we finally derive to leading order in $h_0$
\be
	\bar\varphi_S
	= L^d + \cdots \,, \qquad
	\bar\psi_S 
	=\p_\rho\bar\varphi_S= - \frac{d L^{d}}{h_0 T} + \cdots \,,
\ee
where $\cdots$ only involve terms vanishing in the small $h_0$ limit. Finally,  defining the area operator $A = L^d \O_d$, and substituting these results into \eqref{uncertainty-1}, we obtain
\be\label{final-result}
	\boxed{\big\la (\D A)^2  \big\ra \geq \frac{2\pi G}{d} \la A \ra + \CO(h_0)\,.}
\ee
Importantly, notice that in deriving \eqref{final-result}, we have assumed that our causal diamond is in Minkowski spacetime, as the $\varphi$ given in \eqref{eq:gaussian_null_metric} is diffeomorphic to Minkowski spacetime. It would be interesting to see if the qualitative behavior of area fluctuations change when the causal diamond is embedded in other spacetimes.

The inequality \eqref{final-result} is rather interesting and deserves some further comments. Notice that even though we began by studying the phase space associated to a stretched horizon, the left-hand side of \eqref{final-result} is wholly independent of the stretched horizon $\CH_s$ but only involves the variance of the size of the causal diamond. The reason is because we are taking $h_0 \to 0$, and the variance of the area of the stretched horizon is then to leading order precisely that of the causal diamond. Nevertheless, the usage of the stretched horizon was crucial in our derivation. If we set $\rho_0 = 0$ in the very beginning, this would correspond to using the metric \eqref{metric-cd} rather than \eqref{eq:rho-def}. This was precisely the case considered in \cite{Bub:2024nan}, and the difficulty there is that the coordinate system only describes $\CH^+$, while $\CH^-$ lies outside the coordinate range. Consequently, taking $h_0 \to 0$ is rather delicate, and should be done only at the end of our analysis, since for any infinitesimal $h_0$ we are able to approximate the stretched horizon as $\CH^+ \cup \CH^-$, but setting $h_0 = 0$ in our coordinate system results in the hypersurface $\CH^+$. Thus, the stretched horizon was instrumental in extending our analysis to the past horizon without having to introduce matching conditions by regularizing the geometry near the bifurcating surface. 

Importantly, \eqref{final-result} predicts in the semiclassical regime a lower limit in the uncertainty of the area of a finite causal diamond. The size of quantum fluctuations of the area of a causal diamond are not only sensitive to the UV scale captured by $G$, but also sensitive to the IR scale captured by $\la A \ra$. This is consistent with many of the results derived in the context of black hole backgrounds, such as \cite{Marolf:2003bb, Bousso:2023kdj,Parikh:2024zmu}, where the quantum width of a black hole is computed. We leave a detailed exploration of observable consequences of such results to future work.

\section{Discussion and Future Work}\label{sec:discuss}

In this work, we applied the covariant phase space formalism to study the classical phase space of a stretched horizon living inside a causal diamond in an arbitrary vacuum geometry with neither spin-1 nor spin-2 contributions. By imposing the Raychaudhuri equation, we were able to calculate the constrained symplectic form and obtain the associated Dirac bracket \eqref{bracket-c} between the conjugate modes $\varphi|_{\rho=0}$ and $\psi=\partial_\rho \varphi|_{\rho=0}$. Upon promoting these modes to quantum operators, we found that the product of their uncertainty, given in \eqref{uncertainty-1}, is not simply proportional to $G$, but also depends inversely on the position of the stretched horizon $h_0$ (see Figure~\ref{fig:sh}). Although the uncertainty diverges as $h_0 \rightarrow 0$, suggesting that this limit is singular, we were able to bypass this by angle- and time-averaging $\varphi$ and $\psi$ over the entire stretched horizon in Minkowski spacetime, yielding the perfectly finite result \eqref{final-result} in the $h_0 \to 0$ limit. This states that the minimal variance of area fluctuations in Minkowski spacetime does not only depend on $G$, but also the area itself.

There are a few concrete directions to pursue. As we already mentioned in the introduction, we would like to connect the area fluctuations we have derived in \eqref{final-result} with the modular Hamiltonian fluctuations derived previously in \cite{Verlinde:2019ade,Banks:2021jwj, Verlinde:2022hhs, He:2024vlp}. While the results for area and modular fluctuations looks very similar, they differ by a factor of the transverse dimensions $d$. One key observation is that the modular fluctuations previously studied involve fluctuations corresponding to temperature fluctuations rather than area fluctuations. In our above analysis, we expect such modular fluctuations to correspond to fluctuations in the stretched horizon itself, which is precisely the sector of phase space that we ruled out by imposing $\d h_0 = 0$ in \eqref{h0-constraint}. It would be worthwhile to relax this assumption to study the associated gravitational phase space where we allow the stretched horizon location to fluctuate and relate it to results obtained in the previous literature.

Secondly, in obtaining the minimum area fluctuations of Minkowski spacetime given in \eqref{final-result}, we utilized an angle- and time-averaging of the field $\varphi$ and worked with the simple commutator \eqref{uncertain1}. This is what allowed us to write down such a simple uncertainty relation \eqref{uncertainty-1}. On the other hand, while the commutator \eqref{comm-c} is much more complicated, it has detailed microscopic information on how the fluctuations behave at every point in spacetime along the stretched horizon. We would like to understand how to translate this microscopic information into an uncertainty relation that evolves along the stretched horizon.  We expect this microscopic information to contain the stochastic information of a near-horizon fluid~\cite{Bredberg:2010ky,Bredberg:2011jq}, similar to \cite{Zhang:2023mkf, Bak:2024kzk}, and would like to explore this in more detail in the future.

\section*{Acknowledgements}

We thank Mathew Bub for collaborating in the early stages of this work. We would also like to thank Kwinten Fransen, Laurent Freidel, and Prahar Mitra  for useful conversations. Research at Perimeter Institute is supported in part by the Government of Canada through the Department of Innovation, Science and Economic Development Canada and by the Province of Ontario through the Ministry of Colleges and Universities. L.C. is grateful to K.Z. and T.H. for generously hosting him twice at Caltech, where this project was initiated and brought to completion. T.H. and K.Z. are supported by the Heising-Simons Foundation “Observational Signatures of Quantum Gravity” collaboration grant 2021-2817, the U.S. Department of Energy, Office of Science, Office of High Energy Physics, under Award No. DE-SC0011632, and the Walter Burke Institute for Theoretical Physics. K.Z. is also supported by a Simons Investigator award.

\appendix

\section{General Metric, Raychaudhuri Constraint, and Symplectic Analysis}\label{AppA}

The goal of this appendix is to relate the metric, Raychaudhuri equation, and the pre-sympletic potential we wrote down for our specific analysis with the more general analysis done previously in \cite{Penna:2015gza, Donnay:2015abr, Grumiller:2018scv, Donnay:2019jiz, Chandrasekaran:2021hxc, Adami:2021nnf, Freidel:2022vjq, Freidel:2024emv}. We will in particular be mostly following the notation and results of \cite{Freidel:2024emv}.

\subsection{Line Element}

We begin by demonstrating how our line element \eqref{gengau} in $(d+2)$-dimensional spacetime descends from the general line element describing a stretched horizon. In particular, following \cite{Freidel:2024emv}, the most general line element is given by\footnote{In comparing with \cite{Freidel:2024emv}, we have called the time on the stretched horizon $\tau$ (there called $v$),  and the radial coordinate $\rho$ (there called $r$). Consequently, their function $\rho$ has been renamed $F$ here.} 
\be\label{roma}
\rd s^2=2e^\alpha \big(\rd \tau-\beta_A e^A \big)\big(\rd \rho-F e^\alpha(\rd \tau-\beta_B e^B) \big)+g_{AB} e^A e^B \,,
\ee
where our coordinates are $x^\mu=(\tau,\rho,\vartheta^A)$, with $A=1,\ldots,d$ being the transverse directions, and the frame is related to the coordinates via
\be
e^A=J^A{}_B(\rd \vartheta^B-V^B\rd \tau) \,.
\ee
To obtain \eqref{gengau}, let us define spin-$i$ quantities to be those with $i$ spatial indices $A,B,\dots$. We are interested in fluctuations of the area, which correspond to  spin-$0$ data. This means we can set the spin-$1$ quantities $\beta_A$ and $V^A$ to zero, as they parameterize how vertical and horizontal vectors are twisted due to spin-$1$ contributions. Furthermore, we will work in a coordinate system where $J^A{}_B=\delta^A_B$, so that $e^A=\rd \vartheta^A$. 

Substituting these into \eqref{roma}, we obtain
\be
	\dt s^2 = 2e^\a \,\dt\tau \big( \dt\rho - F(\tau,\rho,\vartheta) e^{\a} \dt\tau \big) + g_{AB}(\tau,\rho,\vartheta)  \dt\vartheta^A \dt\vartheta^B \,.
\ee
Further assuming $\a=0$, and then performing the unimodular decomposition for the angular metric, namely
\be\label{uni-mod}
	g_{AB} = \varphi(\tau,\rho,\vartheta)^{\frac{2}{d}} \bar q_{AB}(\tau,\rho,\vartheta) \,,
\ee
with $\det \bar q_{AB} = 1$, the generic line element reduces to 
\be\label{eq:app-line}
	\dt s^2 = 2 \,\dt\tau \big( \dt\rho - F(\tau,\rho,\vartheta) \dt\tau \big) + \varphi^{\frac{2}{d}}(\tau,\rho,\vartheta) \bar q_{AB}(\tau,\rho,\vartheta) \dt\vartheta^A \dt\vartheta^B \,.
\ee
This is precisely \eqref{gengau} with $\dt\O_d^2 = \bar q_{AB} \dt\vartheta^A \dt\vartheta^B$. Finally, near the stretched horizon, the metric datum $F$ admits the small $\rho$ expansion
\be
	F(\tau,\rho,\vartheta) = h_0(\tau,\vartheta) + \ka(\tau,\vartheta)\rho + \CO(\rho^2) \,.
\ee
Restricting ourselves to the $\rho=0$ stretched horizon, and making the further assumption that $h_0$ is a spacetime constant, we recover the induced metric \eqref{induced} on $\CH_s$, namely
\be\label{eq:app-induced}
	\dt s^2\big|_{\rho = 0} = - 2 h_0  \,\dt\tau^2 + \varphi^{\frac{2}{d}}(\tau,\vartheta) \bar q_{AB}(\tau,\vartheta) \dt\vartheta^A \dt\vartheta^B \,,
\ee
where we have defined $\varphi(\tau,\vartheta) \equiv \varphi(\tau,\rho=0,\vartheta)$ and $\bar q_{AB}(\tau,\vartheta) \equiv \bar q_{AB}(\tau,\rho=0,\vartheta)$ on the stretched horizon. 

\subsection{Raychaudhuri Constraint}

Next, we turn to the generic Raychaudhuri equation on the stretched horizon for the complete line element \eqref{roma}, which is given by
\be\label{ray-gen}
	\CC \equiv \CL_\ell {\cal E}+({\cal E}+\CP)\theta+({\cal D}_A+2\varphi_A){\cal J}^A+{\cal T}^{AB}\sigma_{AB}-\bar\theta\CL_\ell h_0 \,,
\ee
where $\ell$ is the vector tangent to the stretched horizon.  Note that because we fixed $\a = 0$ in the line element \eqref{eq:app-line}, we have $\ell = \p_\tau$. 
Our goal is to derive from \eqref{ray-gen} the Raychaudhuri constraint containing only spin-$0$ data, which is given in \eqref{RayEq}. As before, assuming that no spin-$1$ data contribute, we can set ${\cal J}^A=0$, leading to the simplified Raychaudhuri constraint
\be\label{ray-gen2}
\CC=\p_\tau \CE+({\cal E}+\CP)\theta+{\cal T}^{AB}\sigma_{AB}-\bar\theta \p_\tau h_0 \,.
\ee
The remaining quantities in \eqref{ray-gen2} are
\begin{align}
{\cal E} &= \theta+2h_0\bar\theta \label{cE} \\
{\cal P}&=-\kappa -\frac{(d-1)}{d}(\theta+2h_0\bar\theta) \label{cP}\\
{\cal T}_{AB} &= \sigma_{AB} +2h_0\overline{\sigma}_{AB} \,,\label{cT}
\end{align}
which are interpreted in the membrane paradigm as the energy, the pressure, and the total shear of the putative fluid on the stretched horizon. They are functions of the independent modes $(\theta,\bar\theta,\kappa,\sigma_{AB},\overline{\sigma}_{AB})$, and physically, these modes parametrize $\p_\tau$-expansion, $\p_\rho$-expansion, $\p_\tau$-inaffinity, $\p_\tau$-shear, and $\p_\rho$-shear, respectively. On $\CH_s$, these are explicitly given by
\begin{eqnarray}
	&\th = \p_\tau\log\varphi \big|_{\rho=0} \qquad
	\bar\th = \p_\rho \log \varphi \big|_{\rho=0}  \qquad
	\ka = \p_\rho F \big|_{\rho=0}& \\
	&\s_{AB} = \th_{AB} - \frac{1}{d} \th \varphi^{\frac{2}{d}} \bar q_{AB}  \bigg|_{\rho=0} \qquad
	\overline\s_{AB} = \bar\th_{AB} - \frac{1}{d} \bar\th \varphi^{\frac{2}{d}} \bar q_{AB} \bigg|_{\rho=0} \,,&\label{general-data}
\end{eqnarray}
where we used that the square root of the determinant of the induced degenerate metric is, due to our unimodular decomposition,
\begin{align}
	\sqrt{\det (\varphi^{\frac{2}{d}} \bar q_{AB})} = \sqrt{\varphi^2 \det \bar q_{AB}} = \varphi \,.
\end{align}
Furthermore, note that in \eqref{general-data}, we introduced the expansion tensors (also called second fundamental forms or extrinsic curvature tensors)
\be\label{th-alt}
	\theta_{AB}= \frac12 \CL_{\p_\tau}(\varphi^{\frac{2}{d}} \bar q_{AB}) \,, \qquad 
\bar\theta_{AB}= \frac12 \CL_{\p_\rho}(\varphi^{\frac{2}{d}} \bar q_{AB}) \,.
\ee
The spin-2 data consists of the traceless shears $\s_{AB}$, $\overline\s_{AB}$. In fact, note we can use \eqref{th-alt} to evaluate
\begin{align}
\begin{split}
	\s_{AB} &=  \frac{1}{2} \CL_{\p_\tau} \big( \varphi^{\frac{2}{d}} \bar q_{AB} \big) - \frac{1}{2d} \CL_{\p_\tau} \big( \varphi^{\frac{2}{d}} \bar q_{CD} \big) g^{CD} \varphi^{\frac{2}{d}} \bar q_{AB} \bigg|_{\rho=0} \\
	&=   \frac{1}{2} \varphi^{\frac{2}{d}} \CL_{\p_\tau}\bar q_{AB} - \frac{1}{2d}  \varphi^{\frac{2}{d}} \bar q^{CD} \bar q_{AB} \CL_{\p_\tau} \bar q_{CD} \bigg|_{\rho=0}  \\
	&=   \frac{1}{2} \varphi^{\frac{2}{d}} \CL_{\p_\tau} \bar q_{AB} \bigg|_{\rho=0} \\
	&=  \frac{1}{2} \varphi^{\frac{2}{d}} \p_\tau\bar q_{AB} \bigg|_{\rho=0} \,,
\end{split}
\end{align}
where the second equality uses the inverse of \eqref{uni-mod} and the product rule, and the penultimate equality uses the fact the unimodularity of $\bar q_{AB}$ implies
\begin{align}\label{shear-int}
	0 = \CL_{\p_\tau} \big( \det (\bar q_{AB}) \big) = \det(\bar q_{AB}) \bar q^{AB} \CL_{\p_\tau} \bar q_{AB} =  \bar q^{AB} \CL_{\p_\tau} \bar q_{AB}  \,.
\end{align}
The analogous computation goes through for $\bar\s_{AB}$, thus allowing us to rewrite the shears as 
\be\label{sheshe}
	\sigma_{AB} =\frac{\varphi^{\frac{2}{d}}}{2}\partial_\tau \bar q_{AB} \bigg|_{\rho=0} \,, \qquad
\overline{\sigma}_{AB} =\frac{\varphi^{\frac{2}{d}}}{2}\partial_\rho \bar q_{AB} \bigg|_{\rho=0} \,,
\ee
emphasizing that they are related to the unimodular part of the metric. On the other hand, the spin-0 data consists of the expansion scalars $\th,\bth$, and the inaffinity $\ka$, and it is clear from the definition of the expansion scalars that the breathing mode $\varphi$ is a spin-0 contribution.

As we are only interested in the spin-0 contribution, we also require the shears $\s_{AB}$ and $\overline\s_{AB}$ vanish. Hence, our Raychaudhuri constraint \eqref{ray-gen2} on the stretched horizon becomes
\begin{align}\label{Raysim}
	\CC = \p_\tau\CE + (\CE + \CP)\th\,,
\end{align}
where we used the fact that $h_0$ is constant on the stretched horizon. Finally, substituting \eqref{cE} and \eqref{cP} into this expression, we get
\begin{align}\label{milano}
\CC =  (\p_\tau+\theta)\theta-\mu\theta+2 h_0\left(\p_\tau \bar\theta+\frac{\theta\bar\theta}{d}\right) \,,
\end{align}
where we have defined $\mu=\kappa+\frac{d-1}{d}\theta$ as in \eqref{eq:surface-tension}.\footnote{The quantity $\mu$ was first introduced in \cite{Hopfmuller:2016scf, Hopfmuller:2018fni}, where the authors referred to $\mu$ as the spin-0 momentum. } This is precisely \eqref{RayEq} derived from the line element \eqref{roma}, demonstrating the consistency of our results with the existing literature.

\subsection{Pre-symplectic Potential}

Lastly, we would like to describe the general gravitational phase space on a stretched horizon from a Carrollian viewpoint, starting from the metric \eqref{roma}. Beginning again with  \cite{Freidel:2024emv}, we remove the spin-$1$ data and set $\ell=\partial_\tau$ in the pre-symplectic potential induced by Einstein gravity on the stretched horizon to obtain\footnote{The pre-symplectic potential we are using is analogous to the Dirichlet pre-sympletic potential studied in \cite{Odak:2023pga, Chandrasekaran:2023vzb} for null hypersurfaces, as was thoroughly discussed in \cite{Chandrasekaran:2021hxc}. }
\be
	\Theta = -\frac1{8\pi G}\int_{\CH_s}\ve \left[-\frac12({\cal T}^{AB}+\CP \varphi^{-\frac{2}{d}} \bar q^{AB})\delta (\varphi^{\frac{2}{d}}\bar q_{AB})+\bar\theta \delta h_0\right] + \delta( \cdots) \,,
\ee
where $\ve = \varphi\,\dt\tau\,\dt\O_d$ is the measure on $\CH_s$, and $\d(\cdots)$ indicate terms that are total variations, which do not affect the pre-symplectic form as $\O = \d \Th$ and $\d^2 = 0$. The quantities involved here have all been defined in the previous subsection. In particular, given \eqref{cT}, we know that absence of shears implies ${\cal T}^{AB}=0$, thereby yielding
\be\label{eq:app-symp}
	\Th = - \frac{1}{8\pi G} \int_{\CH_s} \varphi\,\dt\tau\, \dt\O_d \bigg[ -\frac{1}{2}  \CP \varphi^{-\frac{2}{d}} \bar q^{AB}  \d\big(\varphi^{\frac{2}{d}} \bar q_{AB} \big) + \bar\th \delta h_0 \bigg] + \delta(\cdots)   \,.
\ee
Next, note that the unimodularity of $\bar q_{AB}$ implies (cf. \eqref{shear-int})
\be
	0 =  \delta\big(\det(\bar q_{AB}) \big) = \det(\bar q_{AB})\bar q^{AB} \delta \bar q_{AB} = \bar q^{AB}\delta\bar q_{AB} \,.
\ee
Substituting this result into \eqref{eq:app-symp}, we simplify to obtain
\begin{align}
\begin{split}
	\Th &= - \frac{1}{8\pi G} \int_{\CH_s} \varphi\,\dt\tau\, \dt\O_d \big[ - \CP  \varphi^{-1} \d \varphi + \bar\th \delta h_0  \big] + \delta(\cdots)   \\
	&= - \frac{1}{8\pi G} \int_{\CH_s} \dt\tau\, \dt\O_d \bigg[ \bigg( \mu + \frac{2(d-1)}{d} h_0 \bar\th \bigg) \d \varphi + \psi \delta h_0  \bigg] + \delta(\cdots)  \,,
\end{split}
\end{align}
where in the first equality we noted $\bar q_{AB} \bar q^{AB} = d$ and used the chain rule, and in the last equality we used \eqref{cP}, \eqref{eq:surface-tension}, and the definition of $\psi$ in \eqref{meglio}. This is precisely \eqref{thetacan} (up to $\d$-exact terms), and is the starting point of deriving the kinematic symplectic form in Section~\ref{ssec:kinematic}.

\section{Brackets Involving Expansion Scalars} \label{app:grav-phase}

In this appendix, we derive all the Dirac brackets involving the gravitational degrees of freedom $\theta$ and $\bar\theta$. The most straightforward bracket to compute is the one between $\theta$ and itself. Beginning with \eqref{bracket-2} and taking a $\tau$ derivative on both sides, we obtain
\begin{align}\label{bracket-3}
	\big\{ \p_\tau \varphi(\tau,\vartheta) , \th(\tau',\vartheta') \big\} = 0 \quad&\implies\quad \big\{ \th(\tau,\vartheta) , \th(\tau',\vartheta') \big\} = 0 \, ,
\end{align}
where the implication again follows from using the chain rule in conjunction with \eqref{bracket-2}.

Next, we compute the bracket between $\theta$ and $\bar\th$. We have
\begin{align}
\begin{split}
	\big\{\th(\tau , \vartheta), \bar\th( \tau', \vartheta') \big\} &= \bigg\{ \frac{\p_\tau\varphi(\tau,\vartheta)}{\varphi(\tau,\vartheta)} , \bar\theta(\tau', \vartheta') \bigg\} \\
	&= -\frac{\p_\tau \varphi(\tau,\vartheta)}{\varphi(\tau,\vartheta)^2} \big\{ \varphi(\tau,\vartheta), \bar\theta(\tau',\vartheta') \big\}  + \frac{1}{\varphi(\tau,\vartheta)} \p_\tau \big\{ \varphi(\tau,\vartheta) , \bar\theta(\tau',\vartheta') \big\} \\
	&= \p_\tau \bigg( \frac{1}{\varphi(\tau,\vartheta)} D(\tau,\vartheta,\tau',\vartheta') \bigg) \\
	&= - \frac{4\pi G}{h_0} \delta^{(d)}(\vartheta-\vartheta') \p_\tau \bigg[ H(\tau'-\tau) \frac{\th(\tau,\vartheta)}{\varphi(\tau',\vartheta')}\bigg] \,,
\end{split}
\end{align}
where in the second line we used the Leibniz rule for the bracket; in the third line we used \eqref{eq:D-def} and simplified the expression; and in the final line we used \eqref{pizza} and \eqref{bracket-b}. 

Finally, we derive the bracket between $\bar\th$ and itself. To this end, it is convenient to compute the bracket between $\psi$ and itself. Denoting $\varphi \equiv \varphi(\tau,\vartheta)$ and $\varphi' \equiv \varphi(\tau',\vartheta')$ for notational simplicity, and similarly defining $\psi$ and $\psi'$, we note that the second bracket in \eqref{dirac-1} can be written using \eqref{meglio} as
\begin{align}
\begin{split}
	0 &= \bigg\{ \frac{\p_\tau\psi}{\p_\tau\varphi} , \frac{\p_{\tau'}\psi'}{\p_{\tau'}\varphi'} \bigg\} \\  
	&= \frac{1}{\p_\tau\varphi} \p_\tau\p_{\tau'} \big\{ \psi , \psi' \big\} \frac{1}{\p_{\tau'}\varphi'} - \frac{1}{\p_\tau\varphi} \p_\tau\p_{\tau'} \big\{ \psi , \varphi' \big\}\frac{ \p_{\tau'}\psi' }{(\p_{\tau'}\varphi')^2} - \frac{\p_\tau\psi}{(\p_\tau\varphi)^2\p_{\tau'}\varphi'} \p_\tau\p_{\tau'} \big\{ \varphi , \psi' \big\} \,,
\end{split}
\end{align}
where we used the Leibniz and chain rule repeatedly. Isolating the bracket involving $\psi$ and $\psi'$, we obtain using \eqref{bracket-c} and properties of the delta function repeatedly
\begin{align}\label{rho-rho-int}
\begin{split}
	&\frac{1}{\p_\tau\varphi\p_{\tau'}\varphi'} \p_\tau\p_{\tau'} \big\{ \psi , \psi'\big\} \\
	&\quad = \frac{1}{\p_\tau\varphi\p_{\tau'}\varphi'} \bigg[ \p_\tau\p_{\tau'} \big\{ \psi , \varphi' \big\}\frac{ \p_{\tau'}\psi' }{\p_{\tau'}\varphi'} + \frac{\p_\tau\psi}{\p_\tau\varphi} \p_\tau\p_{\tau'} \big\{ \varphi , \psi' \big\}  \bigg] \\
	&\quad = \frac{4\pi G}{h_0} \delta^{(d)}(\vartheta-\vartheta')  \frac{1}{\p_\tau\varphi\p_{\tau'}\varphi'} \bigg[  \p_{\tau'} \big(\delta(\tau-\tau') \p_{\tau'}\varphi' \big)  \frac{ \p_{\tau'}\psi' }{\p_{\tau'}\varphi'} - \frac{\p_\tau\psi}{\p_\tau\varphi} \p_\tau \big(  \delta(\tau'-\tau)\p_\tau\varphi \big) \bigg] \\
	&\quad = \frac{4\pi G}{h_0} \delta^{(d)}(\vartheta-\vartheta')  \bigg[  \p_{\tau} \bigg( \delta(\tau'-\tau)   \frac{\p_{\tau}\psi }{(\p_{\tau}\varphi)^2} \bigg) - \frac{\p_\tau\psi}{(\p_\tau\varphi)^2} \p_\tau  \delta(\tau'-\tau)  \bigg] \\
	&\quad = \frac{4\pi G}{h_0} \delta^{(d)}(\vartheta-\vartheta')  \delta(\tau'-\tau) \p_\tau \bigg(  \frac{\p_{\tau}\psi }{(\p_{\tau}\varphi)^2} \bigg) \,,
\end{split}
\end{align}
which immediately implies
\begin{align}
	\p_\tau\p_{\tau'} \big\{ \psi , \psi'\big\} &= \frac{4\pi G}{h_0} \delta^{(d)}(\vartheta-\vartheta')  \delta(\tau'-\tau) \p_{\tau}\varphi\p_{\tau'}\varphi' \p_\tau \bigg(  \frac{\p_{\tau}\psi }{(\p_{\tau}\varphi)^2} \bigg)  \,.
\end{align}
Integrating this expression twice, we finally obtain
\begin{align}\label{bracket-d}
\begin{split}
	\big\{ \psi , \psi'\big\} &= -\frac{4\pi G}{h_0} \delta^{(d)}(\vartheta-\vartheta')  \int_{\tau}^{\tau_+} \dt \tau''\, H(\tau'-\tau'') (\p_{\tau''}\varphi'')^2 \p_{\tau''} \bigg(  \frac{\p_{\tau''}\p_\rho\varphi'' }{(\p_{\tau''}\varphi'')^2} \bigg) \,.
\end{split}
\end{align}

We can now use this to compute the bracket between $\bar\th$ and itself. First, we compute 
\begin{align}\label{bracket-e}
\begin{split}
	\big\{\bar\th(\tau,\vartheta) , \psi(\tau',\vartheta') \big\} &= \bigg\{ \frac{\psi}{\varphi} , \psi' \bigg\} \\
	&= \frac{1}{\varphi} \big\{ \psi , \psi' \big\} - \frac{\psi}{\varphi^2} \big\{ \varphi , \psi' \big\}  \\
	&= -\frac{4\pi G}{h_0} \delta^{(d)}(\vartheta-\vartheta')  \int_{\tau}^{\tau_+} \dt \tau''\, H(\tau'-\tau'') (\p_{\tau''}\varphi'')^2 \p_{\tau''} \bigg(  \frac{\p_{\tau''}\psi'' }{(\p_{\tau''}\varphi'')^2} \bigg) \\
	&\qquad + \frac{4\pi G}{h_0} \delta^{(d)}(\vartheta-\vartheta') H(\tau' -\tau) \frac{\p_\tau\varphi \ \psi}{\varphi^2} \,,
\end{split}
\end{align}
where we used \eqref{bracket-c} and \eqref{bracket-d}. This in turn implies
\begin{align}\label{bar-theta-bracket}
\begin{split}
	\big\{\bar\th(\tau,\vartheta) , \bar\th(\tau',\vartheta') \big\} &= \bigg\{ \bth(\tau,\vartheta) , \frac{\psi(\tau',\vartheta')}{\varphi(\tau',\vartheta')} \bigg\} \\
	&= \big\{ \bth(\tau,\vartheta) , \psi(\tau',\vartheta') \big\} \frac{1}{\varphi(\tau',\vartheta')} - \big\{ \bth(\tau,\vartheta) , \varphi(\tau',\vartheta') \big\} \frac{\psi(\tau',\vartheta')}{\varphi(\tau',\vartheta')^2} \\
	&=  -\frac{4\pi G}{h_0} \delta^{(d)}(\vartheta-\vartheta') \frac{1}{\varphi'}  \int_{\tau}^{\tau_+} \dt \tau''\, H(\tau'-\tau'') (\p_{\tau''}\varphi'')^2 \p_{\tau''} \bigg(  \frac{\p_{\tau''}\psi'' }{(\p_{\tau''}\varphi'')^2} \bigg) \\
	&\qquad + \frac{4\pi G}{h_0} \delta^{(d)}(\vartheta-\vartheta') \bigg( H(\tau' -\tau) \frac{\p_\tau\varphi \ \psi}{\varphi^2\varphi'}  - H(\tau - \tau') \frac{\p_{\tau'}\varphi' \ \psi'}{\varphi'^2\varphi}  \bigg) \,,
\end{split}
\end{align}
where we used \eqref{bracket-b} and \eqref{bracket-e}. Thus, we see that whereas the bracket between $\th$ and itself vanishes by \eqref{bracket-3}, the bracket between $\bar\th$ and itself does not vanish and is given by the rather complicated expression \eqref{bar-theta-bracket}.

\bibliography{VZeffect-bib}{}
\bibliographystyle{utphys}

\end{document}